\documentstyle[12pt]{article}
\begin{document}
\begin{flushright}
Preprint DFPD 95/TH/37\\
hep-th/9506109 \\
June, 1995
\end{flushright}

\vspace{0.5cm}
\begin{center}
{\Large \bf Duality symmetric actions with manifest space--time
symmetries\footnote{This work
was carried out as part of the
European Community Programme
``Gauge Theories, Applied Supersymmetry and
Quantum Gravity'' under contract SC1--CT92--D789, and
supported in part by M.P.I.}
}

\vspace{1cm}
{\bf Paolo Pasti},
{\bf Dmitrij Sorokin\footnote{on leave from Kharkov Institute of
Physics and Technology, Kharkov, 310108, Ukraine.\\~~~~~e--mail:
sorokin@pd.infn.it} and
Mario Tonin\footnote{e--mail: tonin@pd.infn.it}
}

\vspace{0.5cm}
{\it Universit\`a Degli Studi Di Padova,
Dipartimento Di Fisica ``Galileo Galilei''\\
ed INFN, Sezione Di Padova,
Via F. Marzolo, 8, 35131 Padova, Italia}

\vspace{1.cm}
%\vspace{0.3cm}
{\bf Abstract}
\end{center}
We consider a space--time invariant duality symmetric action for
a free Maxwell field and an
$SL(2,{\bf R})\times SO(6,22)$ invariant
effective action describing a low--energy bosonic sector of the
heterotic string compactified on a six--dimensional torus.
The manifest
Lorentz and general coordinate invariant formulation
of the models is achieved by coupling
dual gauge fields to an auxiliary
vector field from an axionic sector of the theory.


\bigskip
\newpage
\section{Introduction}
The understanding of the important role
played by duality in (super)--Yang--Mills theories \cite{mo,h,sw},
supergravity and the theory
of strings \cite{n}--\cite{sen,sch}
 has allowed
one to make new insight into the structure of these theories and to
find deep relationship between their different (dual) versions.
If target space duality (T--duality) and S--duality
(which is the generalization of
the electric--magnetic duality) are the exact symmetries of string
theory \cite{n,f,w}, it is natural to assume that
there should be a version of the
theory where these duality symmetries are manifest.

A T--duality symmetric string action was proposed
by Tseytlin \cite{ts}
and generalized to a case of
the heterotic string by Schwarz and Sen
\cite{ss1}. This required modification of
space--time transformations of fields.
Note, however, that effective supersymmetric field actions which
describe the low--energy behavior of superstrings do have a global
non--compact  symmetry related to T--duality,
while S--duality is only a symmetry of the equations
of motion \cite{roo}--\cite{ss2}.

In \cite{ss2} Schwarz and Sen proposed models for describing
antisymmetric gauge fields in D--dimensional space--time, where
S--duality symmetry was lifted to the level of action, their results
being the generalization of earlier work by Floreanini and Jackiw
\cite{j} and Henneaux and Teitelboim \cite{ht}
who constructed actions
for describing self--dual tensor fields in (4p+2)--dimensional
space--time (p=0,1,..). The general feature of the models
\cite{ts,ss1,ss2,j,ht} is that,
due to the explicit fixing of the time
direction, they loose manifest Lorentz and general coordinate
invariance,
which, however, are replaced by some modified transformations.
This is
an example how duality effects the symmetry structure of the theory.

Anyway, one may try to look for a formulation of duality symmetric
actions in which conventional space--time symmetries
are restored. An
attempt to do this for a duality
symmetric version of Maxwell theory
\cite{ss2} \footnote{Earlier an alternative version of
duality invariant Maxwell action was
considered in \cite{d}} was undertaken in \cite{l}.
Having been inspired by this paper
we proposed in \cite{pst} a manifestly
space--time and duality symmetric
formulation of (supersymmetric) Maxwell theory by enlarging the
Schwarz--Sen model \cite{ss2} with an
auxiliary vector field $u_m(x)$ and
an antisymmetric tensor field $\Lambda_{mn}(x)$ (m=0,1,2,3)
in such a way that upon
solving for equations of motion and gauge fixing additional local
symmetries associated with the
auxiliary fields the model \cite{pst}
could be reduced to that of Schwarz and Sen. The physical nature of
$u_m(x)$ and $\Lambda_{mn}(x)$  was supposed to be a relic of a
gravitational vielbein and an axion--like potential, respectively.

In the present paper we further develop
the duality symmetric model of
\cite{pst}. In particular, we will show
that the model indeed has a
local symmetry (anticipated in \cite{pst})
allowing one to choose
$u_m(x)$ to be a constant unit--norm time--like vector and,
thus, to demonstrate
that the duality symmetric models  \cite{ss2},
with their non--conventional space--time
symmetries, correspond to a
definite gauge choice for auxiliary fields in
corresponding duality symmetric models
with ordinary Lorentz and general
coordinate invariance. Using the example of a model for two abelian
gauge fields in D=1+3,
we shall show that duality between these two fields
arises due to their specific coupling
to a pseudoscalar sector of the
theory through the
field $u_m(x)$, and the latter can be gauge fixed to a constant
time--like vector by use of a local counterpart of a
Peccei--Quinn symmetry of axion models.
Thus, $u_m(x)$ can originate, in fact, from
the pseudoscalar sector of the theory and, at the same time,
carry some properties of a local Lorentz frame field.
In Section 3 we will present a space--time and
$SL(2,{\bf R})\times O(6,22)$ invariant
effective action which describes
a low--energy limit of a toroidally
compactified heterotic string.

\section{Duality symmetric Maxwell action}
Let us start with an action describing
a free Maxwell field $A_m(x)$ and a
pseudoscalar (`axion') field $a(x)$ in D=1+3
Minkowski space:
\begin{equation}\label{1}
S=\int d^4x \left(-{1\over 4}F_{mn}F^{mn}
-{1\over 2}(\partial_m a(x)-u_m(x))(\partial^m a(x)-u^m(x))
-\epsilon^{pqmn}u_p\partial_q\Lambda_{mn} \right),
\end{equation}
where the first term is the ordinary Maxwell Lagrangian with
$F_{mn}=\partial_mA_n(x)-\partial_nA_m(x)$
and the last two terms form
the Lagrangian which, upon solving the equations
of motion for $u_m(x)$
and using a local symmetry of (\ref{1}) under transformations
\begin{equation}\label{2}
\delta a(x)=\varphi(x),\qquad \delta u_m(x)=\partial_m\varphi(x),
\end{equation}
produces (see, for the details \cite{o,bo,nt,k,q}
and references therein)
the free Lagrangian for the scalar
field $a(x)$
\begin{equation}\label{3}
L=-{1\over 2}\partial_ma(x)\partial^ma(x),
\end{equation}
or its dual
\begin{equation}\label{4}
L={1\over{3!}}\partial_{[m}\Lambda_{np]}\partial^{[m}\Lambda^{np]},
\end{equation}
with the duality relation between $a(x)$ and
the  antisymmetric field
$\Lambda_{np}$ to be
\begin{equation}\label{5}
\partial^la(x)=\epsilon^{lmnp}\partial_m\Lambda_{np}.
\end{equation}
The role of $u_m(x)$ in (\ref{1}) is to be the gauge field of the
symmetry (\ref{2}) and to ensure (\ref{5}) on the
mass shell.

Besides (\ref{2})  the action (\ref{1}) is invariant
under the abelian gauge transformations of $A_m(x)$ and
$\Lambda_{mn}(x)$:
\begin{equation}\label{6}
\delta A_m(x)=\partial_m b(x), \qquad
\delta\Lambda_{mn}=\partial_{[m}b_{n]}(x).
\end{equation}

Note that without violating local symmetry (\ref{2})
one can couple fields from the pseudoscalar sector of
(\ref{1}) to the gauge field in an axion--like fashion:
\begin{equation}\label{ax}
S_{int}=-\int d^4x(\partial_m a - u_m)\epsilon^{mnlp}A_nF_{lp}.
\end{equation}
The sum of (\ref{1}) and (\ref{ax}) is explicitly invariant under
(\ref{2}). It is also invariant under
the gauge transformations (\ref{6})
if one requires that the variation of $\Lambda_{mn}$ acquires the
contribution $\delta_{b}\Lambda_{mn}=b(x)F_{mn}$.
Eliminating either $u_m$ and $\Lambda_{mn}$
or $u_m$ and $a(x)$ by use of
equations of motions and local symmetries
one gets the dual versions of the
axion theory \cite{bo,nt,k,q}.

Now the question arises whether
it is possible to replace (\ref{1})
with an action which would be duality
symmetric in the Maxwell field
sector, still possess ordinary
Lorentz invariance and be equivalent
(at least classically) to the action (\ref{1}).
The answer turns out to be positive \cite{pst}.

{}From \cite{ss2} we learn that for making the
electric--magnetic duality
manifest at the level of action one has
to double the number of abelian
fields (introducing $A^\alpha_m(x)$ ($\alpha$=1,2)) and construct a
duality symmetric action in such a way
that equations of motion obtained
from this action lead to the vanishing of the self--dual tensor
\begin{equation}\label{8}
{\cal F}^\alpha_{mn}={\cal L}^{\alpha\beta}F^\beta_{mn}-F^{*\alpha}_{mn}=
{1\over 2}\epsilon_{mnpq}{\cal L}^{\alpha\beta}{\cal F}^{\beta pq},
\end{equation}
where ${\cal L}^{\alpha\beta}=-{\cal L}^{\beta\alpha}$ (${\cal L}^{12}=1$),
$F^{*\alpha}_{mn}={1\over 2}\epsilon_{mnlp}F^{lp\alpha}$.

When
\begin{equation}\label{9}
{\cal F}^\alpha_{mn}=0
\end{equation}
one of the abelian fields becomes completely
determined through another
one, and for the latter we get the free Maxwell equations by
differentiating (\ref{9}). Then the duality (\ref{9}) between the two
gauge fields reduces to the duality between the electric and magnetic
strength of the gauge field which has been chosen to be independent.

The duality symmetric action proposed in \cite{ss2}, which gives
(\ref{9}), has the following form:
\begin{equation}\label{10}
S=-{1\over 2}\int d^4x(B^{i\alpha}{\cal L}^{\alpha\beta}E^\beta_i
+B^{i\alpha}B_{i}^\alpha),
\end{equation}
where
\begin{equation}\label{eb}
E^\alpha_i=F^\alpha_{0i}=\partial_0A^\alpha_i-\partial_iA^\alpha_0,
\qquad
B^{i\alpha}={1\over 2}\varepsilon^{ijk}F^\alpha_{jk}=
\varepsilon^{ijk}\partial_jA^\alpha_k,
\end{equation}
and $i,j,k = 1,2,3$ are spacial indices.
The action (\ref{10}) is invariant under modified space--time
transformations of $A^\alpha_i$ (in the gauge $A^\alpha_0=0$):
\begin{equation}\label{st}
\delta A^\alpha_i=x^0v^k\partial_kA^\alpha_i+v^kx^k\partial_0A^\alpha_i+
v^kx^k{\cal L}^{\alpha\beta}{\cal F}^\beta_{0i},
\end{equation}
where the first two terms describe the ordinary Lorentz boosts along
a constant velocity $v^i$ and the third term vanishes
on the mass shell (\ref{9}).

The model constructed this way \cite{ss2} is classically and quantum
mechanically \cite{r} equivalent to the free Maxwell theory.

The covariantization of (\ref{10})
is achieved by coupling the self--dual
tensor (\ref{8}) to the auxiliary field $u_m(x)$
from the pseudoscalar part
of the Lagrangian (\ref{1}) as follows:
$$
S=\int d^4x (-{1\over 8}F^\alpha_{mn}F^{\alpha mn}
+{1\over{4(-u_lu^l)}}u^m{\cal F}^\alpha_{mn}{\cal F}^{\alpha np}u_p
$$
\begin{equation}\label{11}
-{1\over 2}(\partial_m a-u_m)(\partial^m a-u^m)
-\epsilon^{mnpq}u_m\partial_n\Lambda_{pq}).
\end{equation}

Action (\ref{11}) differs from that considered in \cite{pst} at the
following point. In \cite{pst} $u_m(x)$ was required to have the
negative unit norm
\begin{equation}\label{n}
u^2\equiv u_mu^m=-u_0u_0+u_iu_i=-1
\end{equation}
and played the role of a
component of a local Lorentz frame. In (\ref{11}) we weakened the
normalization condition by introducing
the norm of $u_m(x)$ only into
the term containing ${\cal F}^\alpha_{mn}$. We shall demonstrate the
relationship between the two versions of the model later on.

The necessity to introduce the norm of $u_m(x)$ into the ${\cal
F}^\alpha_{mn}$--term is dictated by the
requirement to preserve the local
symmetry (\ref{2}). The action (\ref{11}) is invariant under the
transformations (\ref{2}) provided $A^\alpha_m(x)$ and
$\Lambda_{mn}(x)$ are transformed as follows
\begin{equation}\label{12}
\delta A^\alpha_m=
{{\varphi(x)}\over{u^2}}{\cal L}^{\alpha\beta}{\cal F}^\beta_{mn}u^n,
\qquad
\delta\Lambda_{mn}={{\varphi(x)}\over{(u^2)^2}}{\cal F}^{\alpha r}_{m}
u_r{\cal F}^{\beta s}_nu_s{\cal L}^{\alpha\beta}.
\end{equation}
Then, since the solution to the equation of motion
of $\Lambda_{mn}$, obtained from (\ref{11}), is
\begin{equation}\label{13}
u_m(x)=\partial_m\hat\varphi(x),
\end{equation}
where $\hat\varphi(x)$ is a scalar function, we can use the
transformations (\ref{2})
to put $u_m=\delta^0_m$.\footnote{To escape
singularities we should require the norm of $u_m$ to be nonzero.}
In this gauge the action (\ref{11}) reduces to (\ref{10}), and the
transformation of $A^\alpha_m$ in
(\ref{12}) (with $\varphi=x^iv^i$) is
combined with the corresponding Lorentz transformation producing
(\ref{st}).

We see that $u_m(x)$ plays a double role.
{}From the one hand side it is
the gauge field of local Peccei--Quinn symmetry and from the other
hand it corresponds to a component of a local Lorentz frame.

The action (\ref{11}) has another local symmetry \cite{pst}
(which generalizes that
of (\ref{10}) \cite{ss2}) under the following transformations of
$A^\alpha_m$ and $\Lambda_{mn}$:
\begin{equation}\label{ou}
A^\alpha_m~\rightarrow~A^\alpha_m+u_m\varphi^\alpha(x),
\end{equation}
$$
\Lambda_{mn}~~\rightarrow~~\Lambda_{mn}-{\varphi^\alpha\over{u^2}}{\cal
F}^\alpha_{mp}u^pu_n+
{\varphi^\alpha\over{u^2}}{\cal F}^\alpha_{np}u^pu_m.
$$

This symmetry allows one to reduce the general solution of
the equations of motion of $A^\alpha_m$
\begin{equation}\label{a}
\epsilon^{lmnp}\partial_m(u_n{\cal F}^\alpha_{pr}u^r)=0
\end{equation}
to eq. (\ref{9}) (see \cite{ss2,pst} for the details).
In the gauge
where ${\cal F}^\alpha_{mn}=0$,
the equations of motion of $u_m$ lead to the
same expressions for $a(x)$ and $\Lambda_{mn}$ that follow from
(\ref{1}).

To transit from (\ref{11}) to (\ref{1})
we must solve eqs. (\ref{a}) for
one of the gauge fields in terms of another one and substitute the
solution back to (\ref{11}) \cite{ss2,pst}.
If we skip the kinetic term
for $a(x)$ in (\ref{1}) and (\ref{11}),
$\Lambda_{mn}$ becomes a pure
gauge as well, and the three
actions (\ref{1}), (\ref{10}) and
(\ref{11}) become classically equivalent and describe dynamics of a
single Maxwell field.

Now we shall demonstrate how the action (\ref{11}) is related to the
version considered in \cite{pst}. There the vector field $u_m$ (in
(\ref{11})) was
subjected to the normalization condition
(\ref{n}) (we shall denote the
normalized vector as $\hat u_m$).
This caused a problem of establishing
the explicit invariance of the model under
the transformations (\ref{2}),
(\ref{12}). To ensure this invariance one
should couple (\ref{11}) with
the normalized $\hat u_m$ \cite{pst}
to  scale invariant gravity \cite{des}. Then the
action takes the form (we skip, for simplicity, the kinetic term of
$a(x)$):
$$
S=\int d^4x\sqrt{-g}(-{1\over 8}F^\alpha_{mn}F^{\alpha mn}
+{1\over 4}\hat u^m{\cal F}^\alpha_{mn}{\cal F}^{\alpha np}\hat u_p
$$
\begin{equation}\label{con}
-{1\over{\sqrt{-g}}}\epsilon^{pqmn}\hat u_p\partial_q\Lambda_{mn}
+R\Phi^2+6\partial_m\Phi \partial^m\Phi),
\end{equation}
where $g_{mn}(x)$ is a metric, $g=\det g_{mn}$,
$R(x)$ is the scalar curvature and
$\Phi(x)$ is a conformal scalar field.
The action (\ref{con}) is invariant under
(\ref{2}), (\ref{12}) provided
$g_{mn}(x)$ and $\Phi(x)$ subject to the following scale
transformations:
\begin{equation}\label{sc}
\delta g_{mn}(x)=(\hat u^m\partial_m\varphi)^2 g_{mn}(x), \qquad
\delta \Phi(x)=-(\hat u^m\partial_m\varphi)\Phi(x).
\end{equation}
Making redefinition of the metric and $\hat u_m$ as follows
\begin{equation}\label{red}
g^{mn}(x)~~\rightarrow~~\Phi^2g^{mn}, \qquad
\hat u_m(x)~~\rightarrow~~{1\over\Phi}\hat u_m,
\end{equation}
(which preserves the condition (\ref{n})
\footnote{Note that one can introduce
 (\ref{n}) into (\ref{con}) with a Lagrange
multiplier which transforms under the scale transformations in an
appropriate way}) we can rewrite
(\ref{sc}) in the form which describes coupling the model to the
Einstein gravity:
\begin{equation}\label{e}
S=\int d^4x\sqrt{ g}(-{1\over 8}F^\alpha_{mn}F^{\alpha mn}
+{1\over 4}\hat u^m{\cal F}^\alpha_{mn}{\cal F}^{\alpha
np}\hat u_p
-{1\over{\sqrt{g}}}\Phi\epsilon^{pqmn}\hat u_p\partial_q\Lambda_{mn}
+R ).
\end{equation}
The only place where $\Phi(x)$ is present in (\ref{e}) is the term
$\Phi\epsilon^{pqmn}\hat u_p\partial_q\Lambda_{mn}$.
Then, putting $\Phi(x)\hat
u_m(x)=u_m(x)$, making use of (\ref{n}) and taking the flat limit
we get the action (\ref{11}).

This concludes the establishment of the links
between different versions
of the duality symmetric formulation of free Maxwell theory.

\section{Low--energy effective action
in string theory with manifest
$SL(2,{\bf R})\times O(6,22)$ and space--time symmetry}
In this section we present manifest
space--time invariant generalization
of the $SL(2,{\bf R})\times O(6,22)$ invariant low-energy effective
action \cite{ss2} describing heterotic
string theory compactified on a
six--dimensional torus \cite{ms,sen1}.
To do this we should introduce 28
dual pairs \cite{ko} $A^{\alpha,a}_m$ ($a$=1,...,28)
of abelian gauge fields and couple them to scalar
fields in an $SL(2,{\bf R})\times O(6,22)$ covariant way. This is
achieved by modifying the self--dual tensor (\ref{8}) as follows:
\begin{equation}\label{sdm}
{\cal F}^{\alpha,a}_{mn}={\cal L}^{\alpha\beta}L^{ab}F^{\beta,b}_{mn}
-({\cal L}^T{\cal {ML}})^{\alpha\beta}(L^TML)^{ab}F^{*\beta,b}_{mn}
\equiv {\sqrt{-g}\over 2}
(-{\cal L}^T{\cal M})^{\alpha\beta}(L^TM)^{ab}
\epsilon_{mnpq}{\cal F}^{\beta,b~pq},
\end{equation}
where $2\times 2$ matrix valued scalar field
\begin{equation}\label{calm}
{\cal M} =
{1\over \lambda_2(x)}\pmatrix{1 & \lambda_1(x)\cr \lambda_1(x) &
\lambda^2_1+\lambda^2_2\cr},
\end{equation}
satisfies the following conditions:
\begin{equation}\label{co}
{\cal M}^T={\cal M}, \qquad {\cal{MLM}}^T={\cal L}.
\end{equation}
${\cal M}$, $\cal L$ and $A^{\alpha,a}_m$
transform under the global
$SL(2,{\bf R})$ transformations $\omega$
as follows:
\begin{equation}\label{om}
{\cal M}~~\rightarrow~~\omega^T{\cal M}\omega \qquad \omega{\cal
L}\omega^T={\cal L} \qquad A_m=\omega^TA_m.
\end{equation}
The $28\times 28$ matrix--valued scalar field
$M(x)$ satisfies the conditions
\begin{equation}\label{m}
M^T=M,\qquad M^TLM=L,
\end{equation}
where
\begin{equation}\label{l}
L=\pmatrix{ 0 & I_6 & 0\cr I_6 & 0 & 0\cr 0 & 0 & -I_{16}}.
\end{equation}
Under a global $O(6,22)$ rotation $M$, $L$ and $A^{\alpha,a}_m$
transform as follows:
\begin{equation}\label{Om}
M~~\rightarrow~~\Omega^TM\Omega, \qquad \Omega^TL\Omega=L \qquad
A_m~~\rightarrow~~\Omega^TA_m .
\end{equation}
(see ref. \cite{ss2} for the details).
The transformation law of the self--dual tensor (\ref{sdm}) under
$SL(2,{\bf R})\times O(6,22)$ is
\begin{equation}\label{set}
{\cal F}_{mn}~~\rightarrow~~\omega^{-1}\Omega^{-1}{\cal F}_{mn}.
\end{equation}
Using the properties of the fields described above one may convince
oneself that the following general coordinate invariant action has
$SL(2,{\bf R})\times O(6,22)$ symmetry:
$$
S=\int d^4x\sqrt{-g}(-{1\over 8}F^{\alpha,a}_{mn}
({\cal L}^T{\cal {ML}})^{\alpha\beta}(L^TML)^{ab}F^{\beta,b~ mn}
+{1\over {4u^2}}u^m{\cal F}^{\alpha,a}_{mn}
{\cal M}^{\alpha\beta}M^{ab}
{\cal F}^{\beta,b~ np}u_p
$$
\begin{equation}\label{a1}
-{1\over 4}g^{mn}tr(\partial_m{\cal {ML}}\partial_n{\cal{ML}})
+{1\over 8}g^{mn}Tr(\partial_mML\partial_nML)
-{1\over{\sqrt{g}}}\epsilon^{pqmn}u_p\partial_q\Lambda_{mn}
+R ).
\end{equation}
It can be rewritten in a simpler form
$$
S=\int d^4x\sqrt{-g}(
{1\over {2u^2}}u^mF^{*\alpha,a}_{mn}
{\cal F}^{\alpha,a~ np}u_p-
{1\over{\sqrt{-g}}}\epsilon^{pqmn}u_p\partial_q\Lambda_{mn}+R
$$
\begin{equation}\label{a2}
-{1\over 4}g^{mn}tr(\partial_m{\cal {ML}}\partial_n{\cal{ML}})
+{1\over 8}g^{mn}Tr(\partial_mML\partial_nML))
\end{equation}
which, upon fixing the gauge
$u_m={1\over{\sqrt{-g^{00}}}}\delta^0_m$,
$\Lambda_{mn}=0$,
directly reduces to the Schwarz--Sen action \cite{ss2}.

Note that we did not add the kinetic term of $a(x)$ (\ref{1}),
(\ref{11}) to the actions (\ref{a1}) and (\ref{a2}).
This is because
we would like to identify $a(x)$ with
$\lambda_1(x)$ which has already
entered the actions (\ref{a1}) and (\ref{a2}) as part of the
dilaton--axion matrix ${\cal M}$ (\ref{calm}).
The coupling of ${\cal M}$ to the gauge fields ensures the
manifest $SL(2,{\bf R})$ symmetry but
brakes local transformations of $\lambda_1(x)$ (or
$a(x)$ in (\ref{2})) down to
the global Peccei--Quinn shifts which form a
subgroup of the global
$SL(2,{\bf R})$ (\ref{om}) \footnote{Note that
the action (\ref{a1}), (\ref{a2}) is still invariant under the local
transformations of $u_m$ (\ref{2}), $A_m$ and $\Lambda_{mn}$
(\ref{12})}.
Such a coupling violates duality between the pseudoscalar field and
$\Lambda_{mn}$ in favor of the former and makes $\Lambda_{mn}$ an
auxiliary field, which can be eliminated from
(\ref{a1}), (\ref{a2}) by
solving for the equations of motion of $u_m$ and $\Lambda_{mn}$, and
substituting $u_m(x)$ back into (\ref{a1}), (\ref{a2}) in the form
$u_m=\partial_m\hat\varphi(x)$ (\ref{13}).

It is tempting to look for a version
of the low--energy effective string
action which would be manifestly duality
symmetric not only in the gauge
sector but in the axion sector (\ref{1}), (\ref{ax}) as well.
 Might it imply a localization of the $SL(2,{\bf R})$ ?

\section{Conclusion and discussion}
We have constructed the space--time
invariant duality symmetric action
for the free Maxwell theory and the $SL(2,{\bf R})\times O(6,22)$
invariant effective action describing
low--energy bosonic sector of the
heterotic string compactified on a 6--dimensional torus.
This has been
achieved by coupling the self--dual stress tensor,
constructed out of the
dual gauge fields, to the auxiliary
vector field from the axionic sector of the theory.

One can add to the bosonic action (\ref{11}) the kinetic term for
neutral fermions:
\begin{equation}\label{free}
S_F=-i\int d^4x\overline{\psi}\gamma^m\partial_m\psi.
\end{equation}
Then the full action becomes supersymmetric
\cite{ss2,pst} under the
following transformations with odd constant parameters
 $\epsilon^\alpha=i\gamma_5{\cal L}^{\alpha\beta}\epsilon^\beta$:
$$
\delta A^\alpha_m=i{\overline{\psi}}\gamma_m\epsilon^\alpha,
$$
\begin{equation}\label{susy}
\delta{\psi}={1\over 8}F^{\alpha mn}\gamma_m\gamma_n\epsilon^\alpha-
{1\over{4u^2}}{\cal L}^{\alpha\beta}u_p{\cal F}^{\alpha
pm}u^n\gamma_m\gamma_n\epsilon^\beta,
\end{equation}
all other fields being inert
under the supersymmetry transformations.

We see that the supersymmetric transformation law for the fermion
$\psi(x)$ (\ref{susy}) is non--conventional and
reduces to the ordinary
one only on the mass shell (\ref{9}).
This reminds the problem with the
Lorentz transformations (\ref{st}) which we have just solved.
Using the same reasoning as lead us to
introducing $u_m(x)$ one may try
to find a superpartner of $u_m(x)$ whose
presence in the theory gives
rise to a local fermionic symmetry
(being a counterpart of (\ref{2},
\ref{12})) which involves $\psi(x)$
and leads to (\ref{susy}) upon gauge
fixing the local fermionic symmetry.
This construction may arise from
coupling the models discussed above to supergravity, from which, in
fact they originate.

The covariantization procedure for duality symmetric actions
\cite{j,ht,ts,ss2} proposed in \cite{pst} and developed herein is
applicable to abelian tensor fields
in space--time dimensions other than
$D=4$ and may turn out to be useful
for finding new dual versions of $D=10$ supergravity \cite{ss2}.

\bigskip
{\bf Acknowledgements}. The authors are grateful to I. Bandos
for constant interest to this work and helpful comments. D.S. would
also like to thank A. Khoudeir, F. Quevedo and A. Sen for useful
discussion.

\end{document}